# DAMPE silicon tracker on-board data compression algorithm*

DONG Yi-Fan(董亦凡)[1,2]    ZHANG Fei(张飞)[1]    QIAO Rui(乔锐)[1]
PENG Wen-Xi(彭文溪)[1,1)    FAN Rui-Rui(樊瑞睿)[1]    GONG Ke(龚轲)[1]
WU Di(吴帝)[1]    WANG Huan-Yu(王焕玉)[1,1)

1 Institute of High Energy Physics, Chinese Academy of Sciences, Beijing 100049, China
2 University of Chinese Academy of Sciences, Beijing 100049, China

**Abstract:** The Dark Matter Particle Explorer (DAMPE) is an upcoming scientific satellite mission for high energy gamma-ray, electron and cosmic rays detection. The silicon tracker (STK) is a sub detector of the DAMPE payload with an excellent position resolution (readout pitch of 242um), which measures the incident direction of particles, as well as charge. The STK consists 12 layers of Silicon Micro-strip Detector (SMD), equivalent to a total silicon area of $6.5m^2$. The total readout channels of the STK are 73728, which leads to a huge amount of raw data to be dealt. In this paper, we focus on the on-board data compression algorithm and procedure in the STK, which was initially verified by cosmic-ray measurements.

* Supported by the Strategic Priority Research Program on Space Science of Chinese Academy of Sciences (No.XDA040402), and the National Natural Science Foundation of China (No.1111403027).
1) pengwx@ihep.ac.cn, wanghy@ihep.ac.cn

# 1 Introduction

The Dark Matter Particle Explorer (DAMPE) is a space science mission of the Chinese Academy of Sciences. The major scientific objective of DAMPE is to measure electrons and photons in the energy range of 5GeV-10TeV with unprecedented energy resolution (1.5% at 100GeV) in order to explore the possible signatures of dark matter [1]. It will measure the spectra of nuclei up to above 500TeV as well. The DAMPE payload is composed of four sub-detectors as shown in Fig.1, 1) Plastic Scintillator Detector (PSD); 2) Silicon Tracker (STK); 3) BGO calorimeter (BGO) and 4) Neutron Detector (NUD).

The main purpose of the silicon tracker is to measure the incident direction of particles, as well as the charge (Z=1~26). To achieve these goals, a mature technique of silicon micro-strip detectors with high spatial resolution is implemented. Besides, the gamma-ray photons can be also measured by converting them to electron/positron pairs in heavy-Z material(Tungsten) within the STK.

In this paper, we first introduce the system structure of the silicon tracker, the data acquisition architecture, and the requirements for data compression. Then the on-board data compression based on hardware design was presented. At last, an initial verification test was shown.

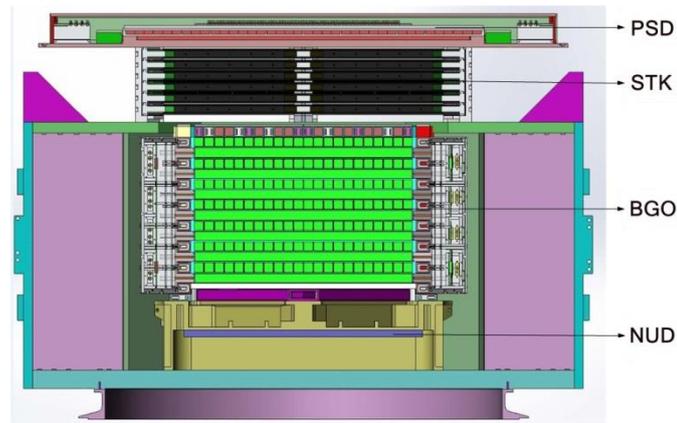

Fig. 1 DAMPE cross section

## 2 The Silicon Tracker

The STK consists of six tracking detector planes, and each plane has two layers (X, Y direction) of silicon detectors with orthogonally oriented strips. Three layers of Tungsten plates with thickness of 1.0 mm are inserted in front of tracking plane 2, 3 and 4 for photon conversion. The silicon detectors are single-sided, AC-coupled, 320um thick, 95mm x 95mm in size, and segmented into 768 strips with a 121um pitch. Four detectors are glued head-on together and wire bonded one after the other along the direction of the strips to form a ladder, and then glued to the Tracker Front-end Hybrid (TFH) where the readout ASICs and associated circuits are integrated. The readout ASIC is the VA140 chip designed by Gamma Medica，a 64-channel, low noise, high dynamic range charge sensitive preamplifier-shaper circuit with simultaneous sample and hold, multiplexed analog readout [2]. On each TFH, there are 6 VA140s to readout 384 strips, only every other strip will be readout. But since analog readout is used as required for charge measurement, the position resolution can be optimized thanks to the charge sharing on floating strips. Each layer has 16 ladders and covers an area about 76cm x 76cm. The Fig.2 shows the arrangement of ladders of the six X-view layers.

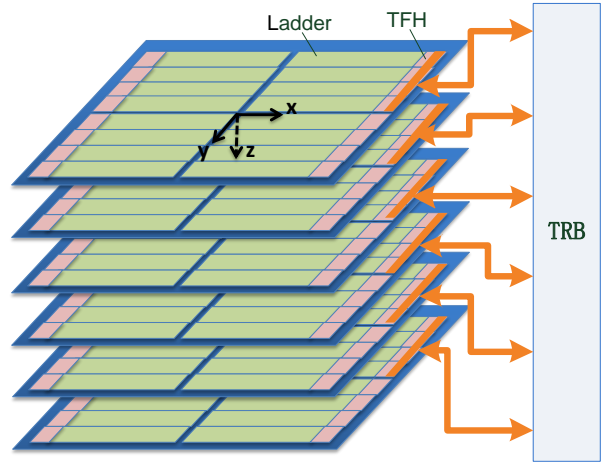

Fig. 2 Ladder arrangement and connection of 6 X-view layers

The ladders are connected via flexible cables to the Tracker Readout Boards(TRBs) surrounding the tracking planes, where the ADCs, monitor circuits, high-voltage generator and control logics are situated. The TRBs are in charge of data acquisition and status monitoring of the silicon tracker. On one hand they receive triggers and configuration commands from payload DAQ, on the other hand, they transfer scientific data and house-keeping data back. There are eight TRBs in all at four vertical sides of the STK, as presented in the Fig.3. Each TRB reads out one quarter ladders of all six layers at the same side, as the Fig.2 shows. The system block diagram is shown in Fig.4.

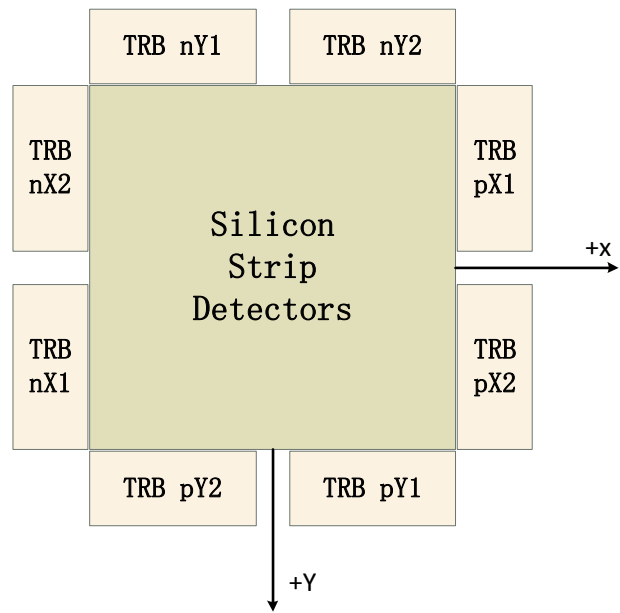

Fig. 3 Top view of the STK

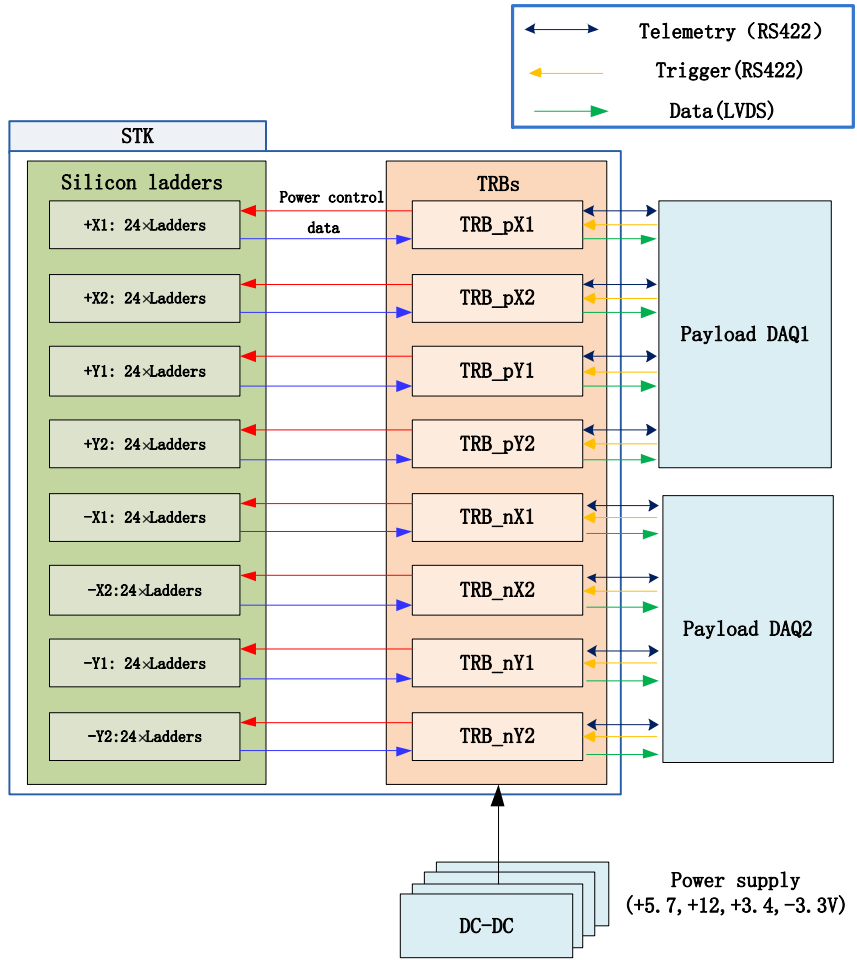

Fig. 4 System block diagram of the STK

# 3 Data compression requirements and challenges

As described above, the STK has 73728 analog readout channels in total. The analog signal of each channel will be digitalized by 12-bit ADC on TRB to possess 2 bytes for each event. According to a physical simulation, the mean trigger rate in-orbit of DAMPE is 50Hz[3], therefore STK will create 637 GB of raw data per day. However the downlink capability of satellite is only 10 GB per day for all the payloads. So it is impossible to download all the raw data of the STK.

On the other hand, in most cases, for each event, only several readout strips are fired due to particle hit, other channels only have noises(zero channel) [4], which should be removed from the real data. This processing method is so called data compression or zero compression.

Actually, there are some similar detectors in space which have data compression on-board. As an example, the AGILE silicon tracker uses the TAA1 readout ASIC which contains a threshold discriminator per read-out channel, and a trigger signal is generated from the OR of all channels of a TAA1 [5]. So that the ASICs which are full of empty channels can be neglected directly before readout, reducing the raw data quite a lot and making the data compression much simpler. However, the DAMPE STK employs the VA140 readout ASIC without self-trigger to meet the strict limit of power consumption. It is noted that the triggers are generated by the BGO calorimeter [6], the STK cannot distinguish fired channels from zero ones before reading out all channels, and as a result, it has to process all channels.

Furthermore, the AGILE has CPUs for data compression but the DAMPE STK has only two FPGAs on each TRB to process about twice number of readout channels than AGILE. Consequently, the DAMPE STK confronts more difficulties compressing the data.

# 4 Data Acquisition Architecture

In the STK, the data compression is done by two FPGAs on each TRB, which also implement all the other logics (controlling readout ASICs, monitoring status and communicating with payload DAQ). Both FPGAs are APA1000 flash FPGA from Actel. Each FPGA takes charge of 12 ladders connected to this TRB. The six VA140 ASICs on each TFH are divided into two groups. In each group, the signals of 192 channels (each ASIC has 64 channels) are output in serial and digitized.

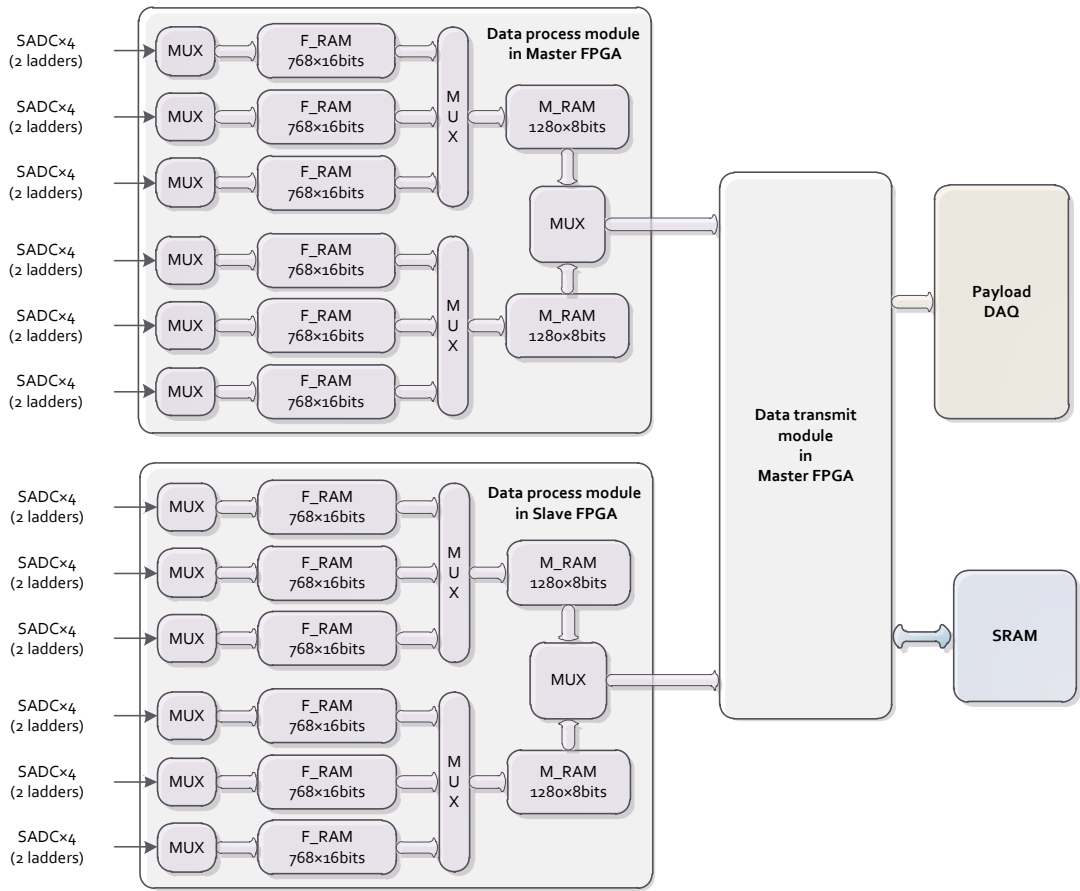

Fig. 5 Data acquisition architecture

At every trigger, each FPGA takes in the data from 24 ADCs in parallel for 192 times, getting raw data of 4608 channels written into the first level cache F_RAM in FPGA. Then pre-processing of data is running, the results are stored in F_RAM as well because of the sources limitation by the FPGA. Every two ladders share a F_RAM unit and each F_RAM unit could be accessed independently so that the pre-processing is in parallel by two ladders. After that, cluster finding is done ladder by ladder, which seeks out the fired channels within some boundary conditions.

Once the step of cluster finding in a ladder is complete, the compressed data will be transferred to the second level cache M_RAM. At last, the data will be packaged in the final format and stored into external SRAM. All these procedures can be finished in 3ms, after that the first and second level cache will be released while the STK is ready for next event.

Generally, the SRAM has the capacity for tens of events in the worst case, the data in SRAM will be transmitted to the payload DAQ at a certain time after every trigger.

# 5 On-board data compression algorithm and procedure

## 5.1 Pre-processing

### 5.1.1 Pedestal subtraction

The pedestal of a channel is its base-level without signal, which is determined by the property of the readout ASIC and the bias. It is defined as:

$$ped_i = \frac{1}{N}\sum_{j=1}^{N} ADC_{ij} \qquad (1)$$

where $ADC_{ij}$ is the digitized value of each channel i in the event j with a random trigger. Fig.6 shows a typical pedestal distribution of a STK ladder.

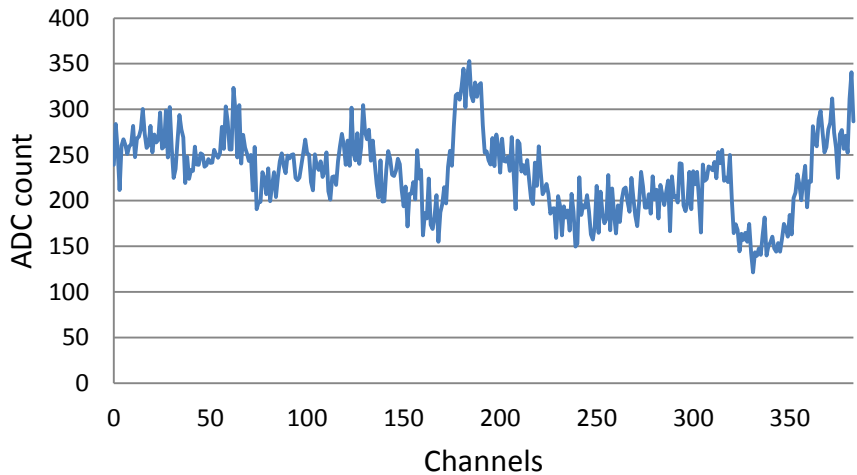

Fig. 6 Pedestal of a ladder

There is a pedestal updating procedure in FPGA to calculate the pedestals on-board using 1024 times of random triggers. It operates twice every orbit cycle and the computed pedestal values are stored in TRB.

During the observation mode of STK, the first step of data compression is to subtract the pedestals from the raw ADC values. The results will first replace the raw ADC values in the F_RAM cache and then act as input in the next step.

## 5.1.2 Common mode noise subtraction

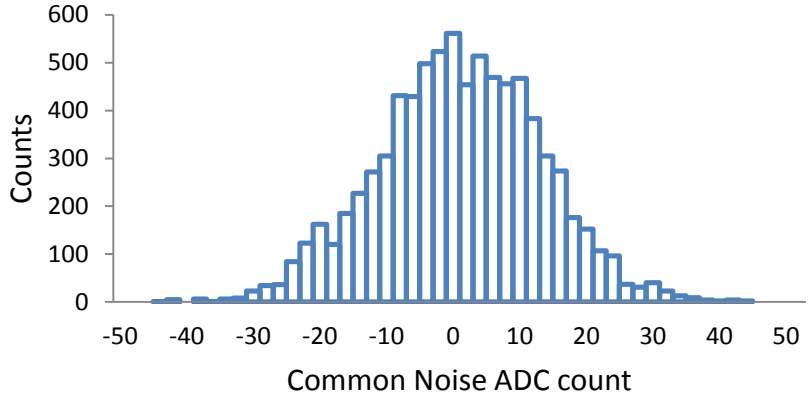

Fig. 7 Common mode noise distribution of a ladder

The common mode noise (CN) is the deviation of all the channels of a readout ASIC at the same time mainly because of the fluctuation on the bias voltage. For each ASIC, the common mode noise of event j is calculated as

$$CN_j = \frac{1}{N_j} \sum_i^{N_j} (ADC_{ij} - ped_i) \quad (2)$$

where $N_j$ is the number of the good strips within the ASIC (noisy or dead strips need to be excluded), Fig.7 shows a typical distribution of the common noise of a ladder. Usually $N_j$ is 64 for the STK, the total channel number of the VA140, but if bad strips exist (as seen in Fig.8 and Fig.9), it is hard to calculate CN in FPGA when $N_j$ is no longer power of 2. As a solution in FPGA, the CN is actually computed with 32 good channels, which is almost the same as with 64 good channels. For the worse case when bad channels are more than 32, the CN can be computed with 16 good channels automatically. But if most of the channels are bad (>48), all channels of the ASIC will be blocked.

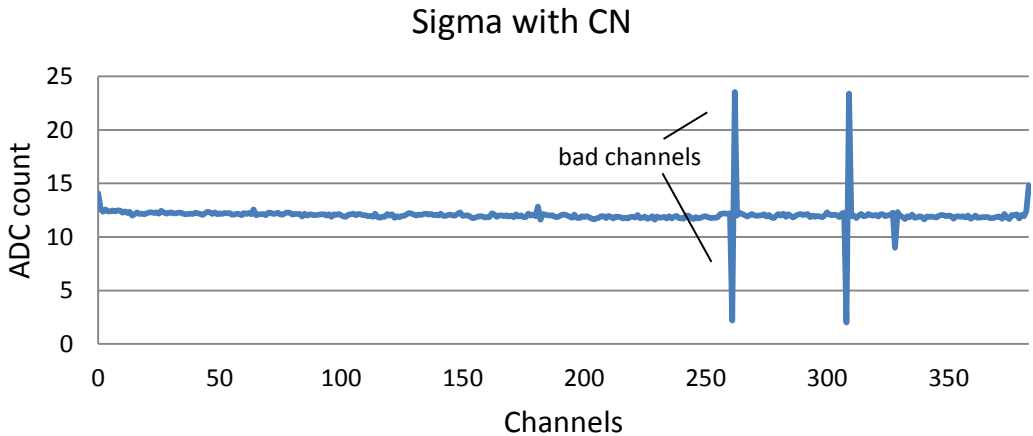

Fig. 8 Sigma with CN

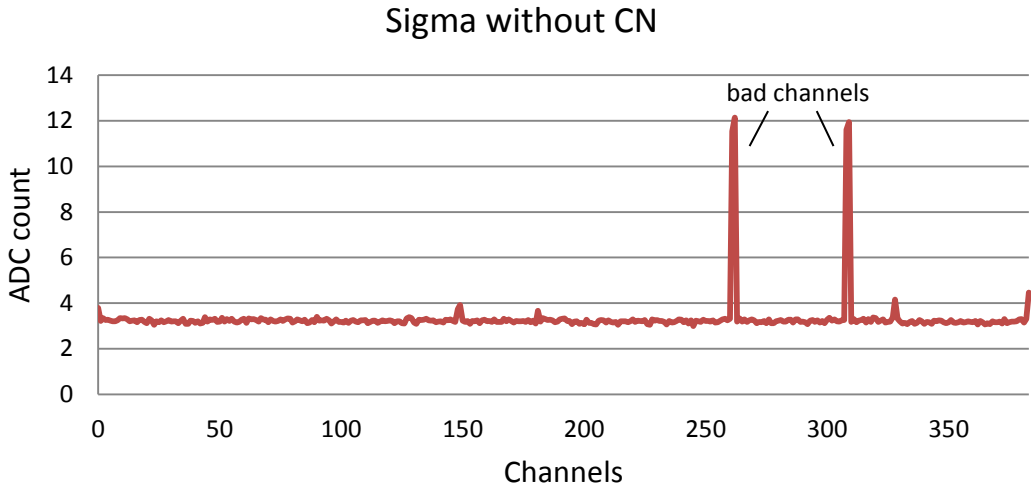

Fig. 9 Sigma without CN

The influence of CN on the sigma value is showed in Fig.8 and Fig.9. A too high CN could make noises confounding signals. In the second step of data pre-processing, the CN will be computed for each event and subtracted, replacing the results in the memory of last step.

## 5.1.3 Cut bad channel

So far the reduced values in the memory are

$$r_{ij} = ADC_{ij} - ped_i - CN_j \quad (3)$$

In the last step of pre-processing, a bad channel cutting is applied to the reduced values by writing them to zero which will be certainly excluded in cluster finding. The bad channels are identified by a bad channel list stored on-board, which could be modified by sending telemetry commands from ground.

In addition, the pedestals of all channels calculated and stored on-board for pedestal cutting are provided with an odd check. If the pedestal of a channel is mistaken because of single event upset (SEU), the reduced value of the channel will also be written to zero in this step because the subtracted pedestal is unreliable.

## 5.2 Cluster finding

A cluster is a group of neighboring strips sharing the charge induced by the energy deposited by the particle. These strips are the fired ones which need to be found out and kept in the compressed data. The cluster finding begins with the traversing of the reduced values in the memory and the seeking of the cluster seed. A cluster seed is a channel with the reduced value $r_{ij}$> $Ts_i$, the $Ts_i$ is the seed threshold of the channel i. When there is a cluster seed, the channels before and behind are checked with an identical fire threshold Tf, until $r_{ij}$< Tf. The whole cluster is from the first channel to the last channel which all have $r_{ij}$>=Tf, with at least one channel having $r_{ij}$>$Ts_i$ within the cluster.

The address of the first channel of the cluster and the reduced value of each channel are saved in the compressed data. Since the channels are continuous, the address of the first channel is sufficient to get the hit information.

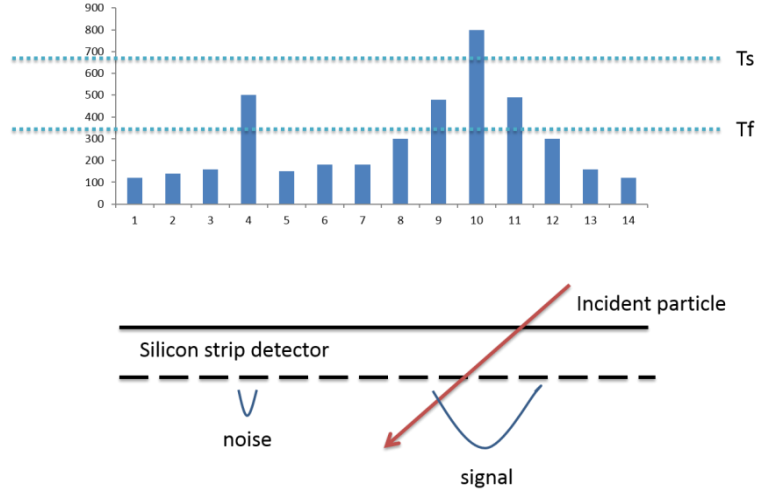

Fig. 10 Cluster finding

In addition to the observation mode with data compression, the STK also has a raw data mode running for in-flight calibration. The sigma of each channel could be computed from the raw data, using

$$\sigma_i = \sqrt{\frac{1}{N}\sum_{j=1}^{N}(ADC_{ij} - ped_i - CN_j)^2} \quad (4)$$

where $ped_i$ and $CN_j$ are computed as noted. The seed thresholds and fire thresholds, which supposed be a multiple of the sigma, are decided offline and sent to the STK to do modification by telemetry commands.

In this algorithm, the data compression ratio is related to the number of clusters per event and the number of strips per cluster. According to the STK physical simulation [4], the average number of clusters per event is 190 and the average number of strips per cluster is 3, thereby the total compressed data per day with 50Hz trigger rate will be 6.3 GB, which is less than 1% of raw data volume.

# 6 Test results with cosmic muons

The on-board data compression procedure has been tested by measuring cosmic muons. The trigger is generated from an additional plastic scintillator on the top of silicon ladder. The same number of raw data events and compressed data events are compared, the results are shown in Fig.11 and Fig.12. The Landau fit is used to obtain the most probable value and the sigma of the MIPs peak, the cluster energy spectrum of raw data and compressed data are agreed quite well which proves that the data compression keeps the proper information.

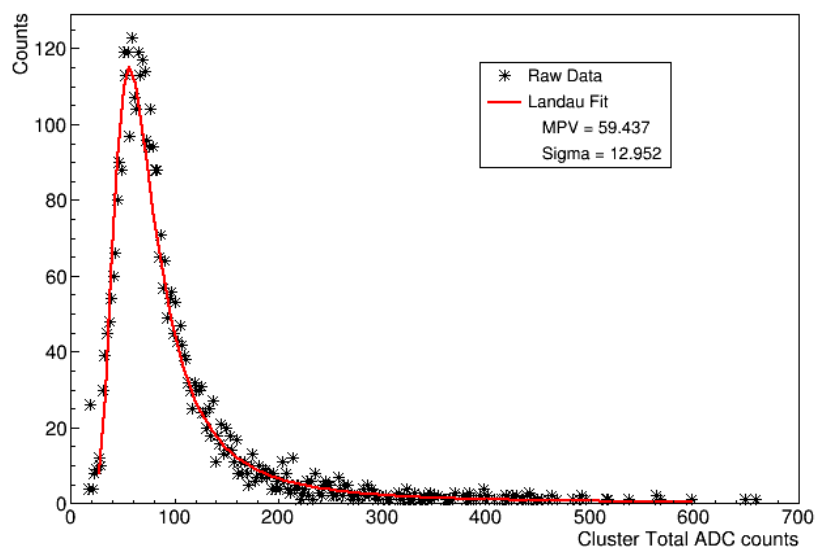

Fig. 11 Cosmics spectrum of raw data

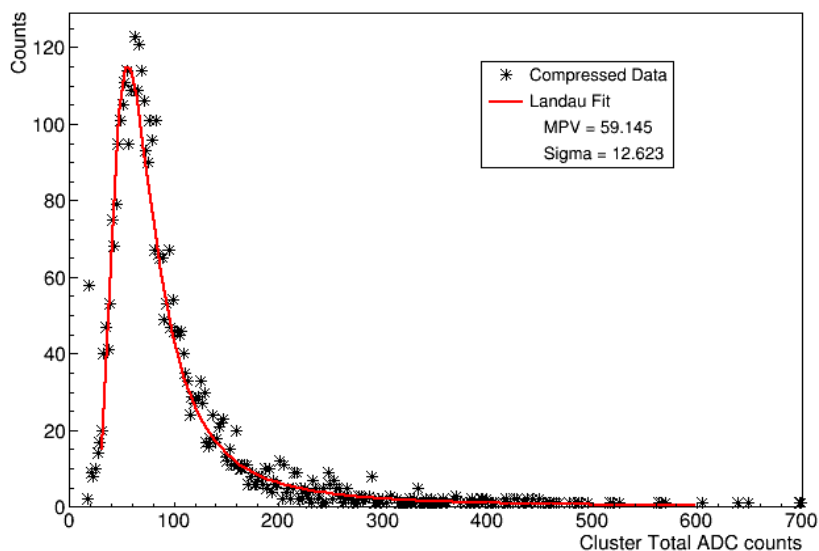

Fig. 12 Cosmics spectrum of compressed data

# 7 Summary

The silicon tracker of the DAMPE uses silicon micro-strip detectors to achieve an excellent position resolution, which results in a huge number of readout channels and thus a huge amount of raw data. According to the downlink limitation of the satellite, on-board data compression is required for the STK, which is challenging to accomplish with restricted on-board resources. This paper presents the data acquisition architecture and the data compression algorithm and procedure of the STK. The initial test results with cosmic muons demonstrate that the on-board data compression of the STK is valid and effective.

# References

[1] CHANG J. Chin. J. Space Sci., 2014, 34(5): 550-557

[2] Gamma Medica-Ideas. VA140 documentation(V0R1) 2011

[3] The trigger and DAQ in DAMPE, DAMPE Workshop June 2013

[4] Wu X, Triggering on (Low Energy) Photons and Data Volume of STK, DAMPE Workshop November 2013

[5] Bulgarelli A, et al. Nucl. Instr. and Meth. A, 2010, 614: 213–226

[6] FENG C Q, et al. Prototype design of DAMPE Calorimeter readout electronics and performance in CERN beam test, 40th COSPAR Scientific Assembly, 2014